\documentclass[preprint,showpacs,preprintnumbers,amsmath,amssymb]{revtex4}

\usepackage{graphicx}
\usepackage{dcolumn}
\usepackage{bm}

\begin{document}

\title{\large Tunneling and bound states in near-edge NbN-$\rm Bi_2Se_3$ junctions fabricated by deposition through a wire shadow mask }

\author{G. Koren}
\email{gkoren@physics.technion.ac.il} \affiliation{Physics
Department, Technion - Israel Institute of Technology Haifa,
32000, ISRAEL} \homepage{http://physics.technion.ac.il/~gkoren}

\date{\today}
\def\bfig {\begin{figure}[tbhp] \centering}
\def\efig {\end{figure}}

\normalsize \baselineskip=8mm  \vspace{15mm}


\begin{abstract}

Transport measurements in thin film junctions of NbN-$\rm Bi_2Se_3$ exhibit tunneling as well as bound state resonances. The junctions are prepared by pulsed laser deposition of a NbN layer through a 25 $\mu$m wide gold wire shadow mask bisecting the wafer into two halves, on a $\rm Bi_2Se_3$ blanket film without further patterning. This results in two independent near-edge junctions connected in series via the 25 $\mu m$ long and 10 mm wide area of the uncapped $\rm Bi_2Se_3$ layer. Conductance spectra measured across the wire masked trench at different locations on the wafer show that some junctions have tunneling behavior with pronounced coherence peaks at $\pm 2\Delta$ where $\Delta \simeq$ 1 meV, while others have zero bias conductance peaks and series of bound states at higher bias. The later can be attributed to zero energy Majorana bound states or to the more conventional Andreev bound states. Based on the present results, we can not distinguish between these two scenarios.

\end{abstract}

\maketitle

\section{Introduction }
\normalsize \baselineskip=6mm  \vspace{6mm}

Topological superconductivity (TOS) is predicted to support zero mode Majorana bound states (MBS) which will possibly be used in future quantum computers \cite{KaneRMP,Kitaev}. In many studies researchers have tried to realize TOS by inducing superconductivity in a topological insulator (TOI) or in a semiconductor with strong spin-orbit interaction, by means of the proximity effect with a conventional s-wave superconductor \cite{Oreg,KorenPRB12,Kouwenhoven,Moti,KorenEPL13,KorenSUST15,CMarcus}. While there is increasing experimental evidence for the presence of MBS in these proximity systems, the subject is still controversial with regard to their actual existence. The present study is an extension of our previous reports on such a proximity system comprised of $\rm Au-Bi_2Se_3-NbN$ junctions \cite{KorenPRB12,KorenEPL13,KorenSUST15}, but here the junctions are prepared in one step by wire shadow masking with no further patterning, thus keeping them pristine and avoiding deterioration by the photolithographic and ion milling processes. In addition to the observation in the present study of tunneling and zero bias conductance peaks (ZBCPs), as already observed in our previous studies \cite{KorenPRB12,KorenEPL13,KorenSUST15}, here we also find sub-gap linear series of non zero bias bound states which can be due to multiple Andreev reflections in the $\rm Bi_2Se_3$ base layer, resulting in Andreev bound states (ABSs). \\

\section{Preparation and characterization of the films and junctions }
\normalsize \baselineskip=6mm  \vspace{6mm}

The NbN and $\rm Bi_2Se_3$ thin films were prepared by laser ablation deposition from a metallic Nb target and from a non stoichiometric pressed pellet with Bi:Se ratio of 1:17 target. While the NbN film was deposited under 30 mTorr of $\rm N_2$ gas flow the $\rm Bi_2Se_3$ film was deposited under vacuum. Both films were deposited at 300 $^0$C heater block temperature, which yields about 250 $^0$C on the surface of the wafer. Due to this low deposition temperature for the cap NbN layer which is required in order to avoid Se loss from the base $\rm Bi_2Se_3$ film, the resulting superconducting transition temperature $\rm T_c$ is only 9.5 K, instead of the normally obtained 12-14 K when deposition is done at 600 $^0$C heater block temperature. High laser fluence on the target was used for the deposition of NbN ($\rm \sim 10\, J/cm^2$), while a much lower fluence was needed for the deposition of the $\rm Bi_2Se_3$ film (\rm $\sim 1\, J/cm^2$) which also helped to reduce the Se loss. Both films were deposited on (100) $\rm SrTiO_3$ (STO) wafers of $\rm 10\times10\, mm^2$ area. X-ray diffraction measurements of single layers of NbN and $\rm Bi_2Se_3$ on (100) STO showed that both grew with preferential crystallographic orientation. The NbN layer, mostly in the cubic phase with a-axis orientation and a=0.433 nm, while the $\rm Bi_2Se_3$ film had the typical hexagonal structure with c-axis orientation normal to the wafer and c=2.88 nm.   \\

The junctions were prepared by first depositing a 60 nm thick blanket film of $\rm Bi_2Se_3$ on (100) STO wafer. Then a straight gold wire of 25 $\mu$m diameter was attached in contact as a shadow mask to the wafer by two silver paste dots on its two edges, in such a way that it divided the wafer into two 10$\times$5 mm$^2$ halves. This was followed by the deposition of a 60 nm thick NbN layer, yielding an NbN/$\rm Bi_2Se_3$ bilayer with a 25 $\rm \mu m \times 10$ mm trench in the NbN layer along the bisector of the sample. Transport measurements were carried out by the use of an array of 40 gold coated spring loaded spherical tips for the 4-probe dc measurements on ten different locations on the wafer, with silver paste contact pads. Each of the four sets of ten contacts was equidistantly separated from one another by 1 mm and the bias current was flown across the trench. Magnetic fields of up to 4 T were applied normal to the wafer along the c-axis direction of the $\rm Bi_2Se_3$ layer. \\

\section{Results and discussion}

Fig. 1 depicts resistance vs temperature at zero magnetic field at four different locations on the wafer which we term junctions J2, J4, J6 and J7. The normal state resistance is mostly that of the two layers connected in parallel, and is quite constant vs T down to the transition temperature at T=$\rm T_c$(onset) of about 9.5 K. Below it R drops quite sharply for all junctions, then two kinds of resistive proximity tails are developed (see left side inset with R on a logarithmic scale), but the R values of all junctions remains resistive down to 2 K. For the J2 and J4 junctions the residual resistance at 1.9 K is 0.24 $\Omega$ and is due to the serial resistance of the normal 25 $\mu$m long $\rm Bi_2Se_3$ under the trench, but for J6 and J7 the resistance is of about 3.1 $\Omega$, and its temperature dependence is shown in the right hand side inset. The later shows increased resistance with decreasing temperature down to 5 K, but then an opposite behavior of decreasing R down to 2 K. The increasing part can be due to the insulating behavior of $\rm Bi_2Se_3$ at low temperature \cite{Butch} as well as to flux flow resistance in the NbN layer, while the decreasing resistance part seems to result from flux pinning. The fact that two kinds of junctions with very different resistance at low temperature are observed in the present sample, means that the cap NbN layer does not completely short the different locations on the wafer, and that some weak links must exist to separate between these two sets of junctions. A similar variation of resistance on the same wafer was also observed in microbridges patterned in trilayers of $\rm Bi_2Se_3$ / $\rm SrRuO_3$ / $\rm YBa_2Cu_3O_x$ \cite{KorenSUST18}. We thus expect here different types of conductance behavior from the two kinds of junctions. \\

\begin{figure} \hspace{-20mm}
\includegraphics[height=9cm,width=11cm]{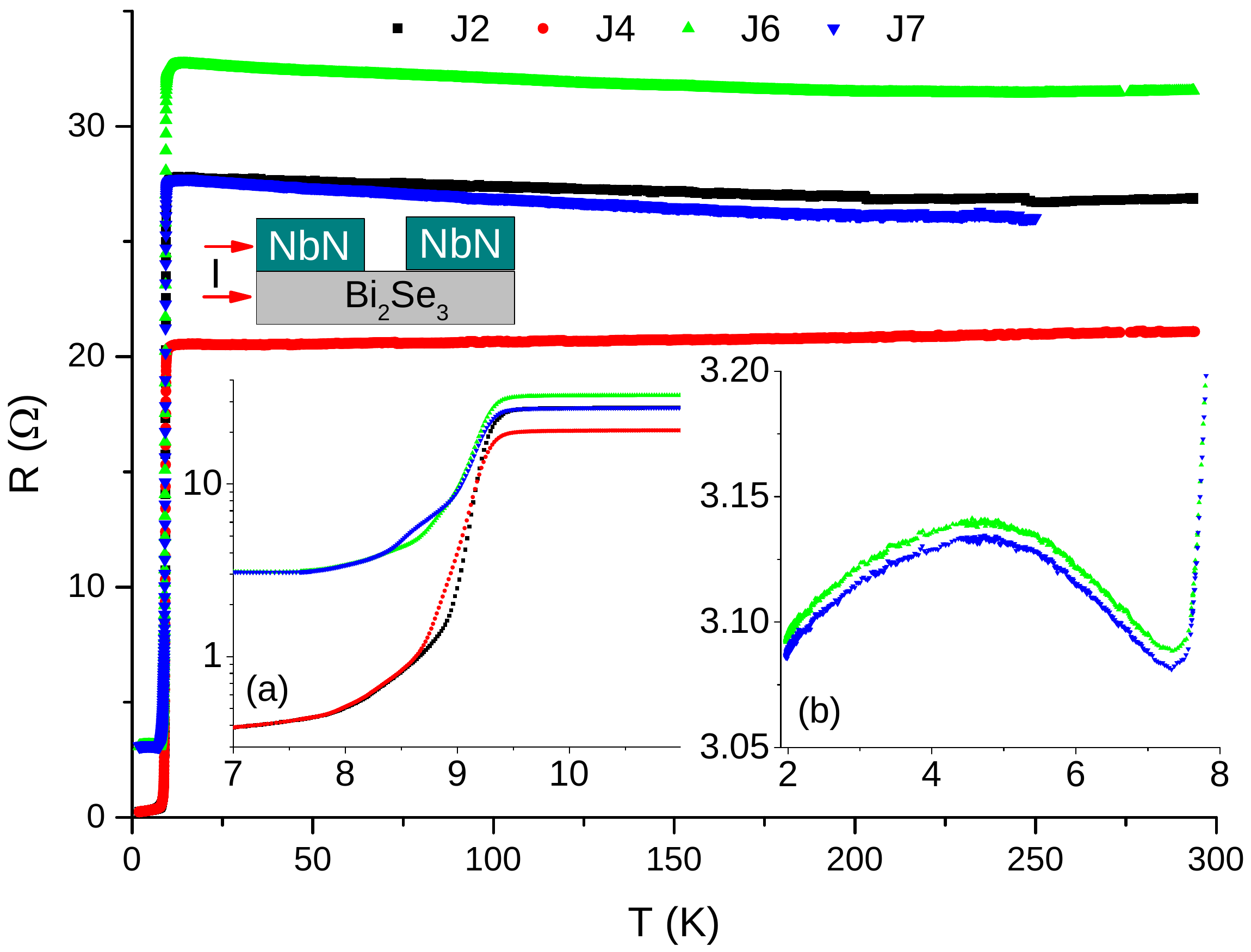}
\vspace{-0mm} \caption{\label{fig:epsart} (Color online) Resistance versus temperature across the present wire-masked junctions J2, J4, J6 and J7 measured at four different locations on the $10\times 10$ mm$^2$  wafer area. The insets show zoom-ins on the superconducting transition regime, and in the schematic drawing each layer is 60 nm thick. }
\end{figure}

\begin{figure} \hspace{-20mm}
\includegraphics[height=9cm,width=11cm]{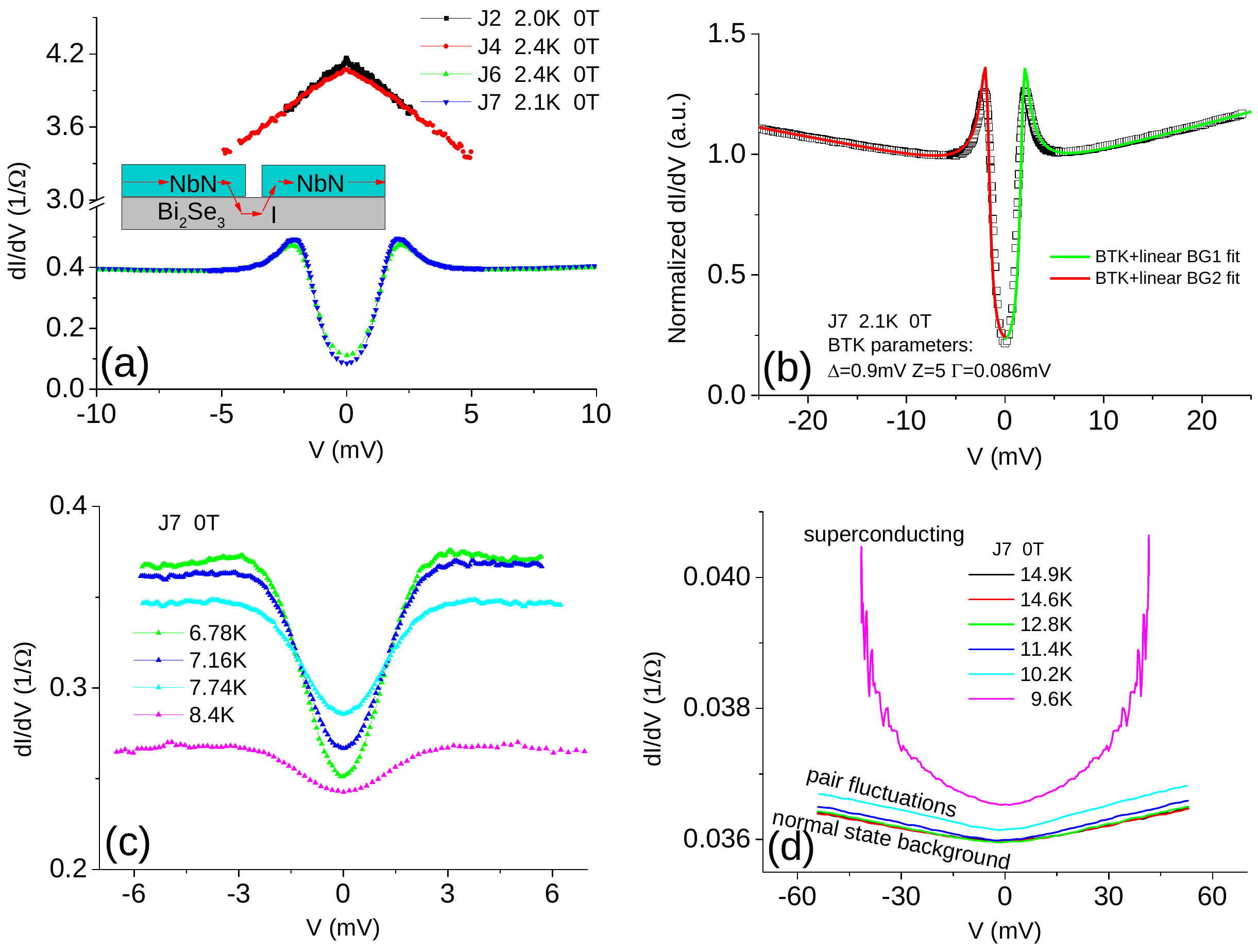}
\vspace{-0mm} \caption{\label{fig:epsart} (Color online) Panel (a) depicts the conductance spectra of all four junctions of Fig. 1 at 0 T and low temperatures. In panel (b) the normalized spectrum of the J7 junction at 2.1 K is fitted using the BTK model of Ref. \cite{BTK}, with two slightly different linear backgrounds added to the positive and negative parts of the spectrum.  Spectra of the J7 junction at different temperatures below and above the transition temperature at 9.5 K are plotted in (c) and (d), respectively.  }
\end{figure}

The measured conductance spectra at low temperatures and zero field of all junctions of Fig. 1 are plotted in Fig. 2 (a), with a broken ordinate scale. One can easily see that the J6 and J7 junctions have a typical tunneling spectra with low zero bias conductance and two prominent coherence peaks. The spectra of the J2 and J4 junctions are more Andreev like, but clearly a background has to be subtracted before characterizing them further. First we shall focus on the J7 junction. In Fig. 2 (b) we show s-wave BTK fits \cite{BTK} to the J7 spectrum of (a). Since we are dealing here with two tunneling junctions connected in series (see the current flow in the schematic drawing in Fig. 2 (a)), the coherence peaks appear at $2\Delta$. The tunneling behavior also indicates a high Z value tunneling barrier. The thermal broadening $\Gamma$ of the spectra was taken as kT at 2.1 K. In addition due to the slightly asymmetric spectrum, two different linear backgrounds were added to the fits. The resulting fit parameters are given in this figure, and one can see an excellent fit to the measured data. Fig. 2 (c) and (d) show conductance spectra of J7 at 0 T versus temperature below and above $\rm T_c$, respectively. Above 10 K one can see the almost linear normal background contribution to the conductance. At 9.6 K superconductivity sets in, and below it the superconducting gap becomes deeper with decreasing temperature and the coherence peaks more pronounced. Also the normal state conductance above 3 mV increases with decreasing temperature. This seems to originate in more percolative superconducting paths in the NbN layer between the voltage contacts with decreasing temperature close to $\rm T_c$. This is consistent with the previously described scenario in which the NbN layer contains weak-links, thus allowing for different behaviors at different locations on the wafer. All these characteristics are typical of tunneling junctions with low transparency (high Z value). \\

\begin{figure} \hspace{-20mm}
\includegraphics[height=9cm,width=13cm]{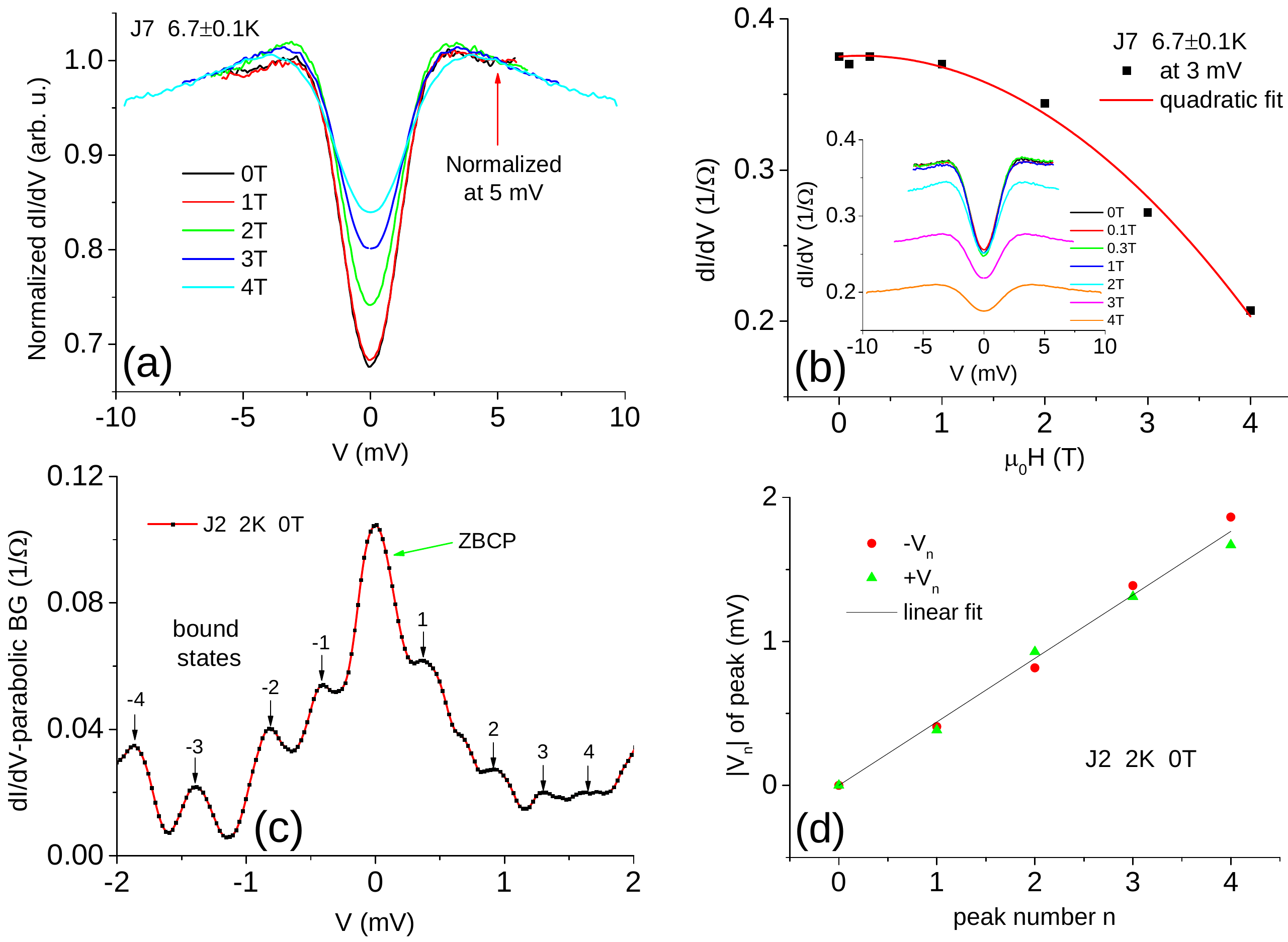}
\vspace{-0mm} \caption{\label{fig:epsart} (Color online) Normalized conductance spectra of the J7 junction at 6.7$\pm$0.1 K under different magnetic fields normal to the wafer are plotted in (a). The un-normalized data is shown in the inset of (b) and the data in (a) is normalized at 5 mV. In (b) the the conductance of the C7 junction at 3 mV is plotted versus magnetic field, together with a quadratic fit which indicates flux flow resistivity \cite{Tinkham}. In (c) the conductance spectrum minus a parabolic background of junction J2 at 0 T and 2 K is plotted. It exhibits a ZBCP and series of bound state peaks whose voltage bias values $\rm |V_n|$ are plotted vs the peak number n in (d).         }
\end{figure}

The normalized conductance spectra of J7 at 6.7$\pm$0.1 K are shown in Fig. 3 (a) under various magnetic fields normal to the wafer. The un-normalized data is shown in the inset to Fig. 3 (b). One can see that the gap depth decreases with increasing field (a), while the zero bias conductance decreases for fields higher than 2 T (see the inset of (b)). The main panel of (b) depicts the conductance at 3 mV which fits a parabolic behavior vs field. This is typical of flux flow \cite{Tinkham}, thus this phenomenon in J7 is well established. We now turn to deal with the J2 conductance spectrum of Fig. 2 (a). It seems to have a significant parabolic  background and we  note that unlike the linear background in Fig. 2 (b) which extends to $\pm$25 mV, here the voltage range is much smaller, a mere $\pm$2 mV, therefore a quadratic background is more appropriate. By subtracting such background from the conductance spectrum of J2 we obtain the result of Fig. 3 (c). Although the signal is quite small now, one can see a clear zero bias conductance peak (ZBCP) together with non-zero bias equidistant peaks distributed symmetrically at positive and negative bias values. In (d) we plot the voltages $\rm |V_n|$ of the peaks versus n, the peak number, and find that they follow a linear behavior. We attribute the ZBCP to either MBS or ABS. For this to occur, an unconventional superconductivity has to be induced at the interface of the topological and superconductor layers, such as in the Eu2 symmetry, the $\rm p_x+ip_y$ p-wave symmetry, or other order parameters with nodes as  observed previously \cite{KorenPRB12,KorenEPL13,FuBerg}. The non zero bias series of peaks can be due to ABS created by multiple Andreev reflections normal to the wafer at the base $\rm Bi_2Se_3$ layer. Similar series of equidistant peaks was found in a theoretical analysis of $\rm N_1-I-N_2-S$ junctions with d+is order parameter where $\Delta_s/\Delta_d=1/3$  in the superconductor S, $\rm N_2$ is the proximity layer, I is a delta function barrier of strength Z as in the BTK model \cite{BTK}, and $\rm N_1$ is a normal metal layer (see Fig. A4 of the Appendix) \cite{Lubimova}. A similar type of junction was discussed in Ref. \cite{N1IN2S} for a pure d-wave superconductor.\\

\section{Conclusions}

Two types of near-edge NbN-$\rm Bi_2Se_3$ thin film junctions connected in series via a short $\rm Bi_2Se_3$ segment were fabricated by a wire shadow mask with no further patterning. Measurements at different locations on the wafer show that at low temperatures some junctions were an order of magnitude more resistive  than others. The former exhibit classical tunneling conductance spectra with low interface transparency, while the later show ZBCP and a linear series of bound states. The ZBCP could originate in MBS as well as ABS, but the present data does not allow for separation between the two.\\


\newpage

\begin{figure} \hspace{-20mm}
\renewcommand{\figurename}{Fig. A\hspace{-1.5mm}}
\includegraphics[height=9cm,width=13cm]{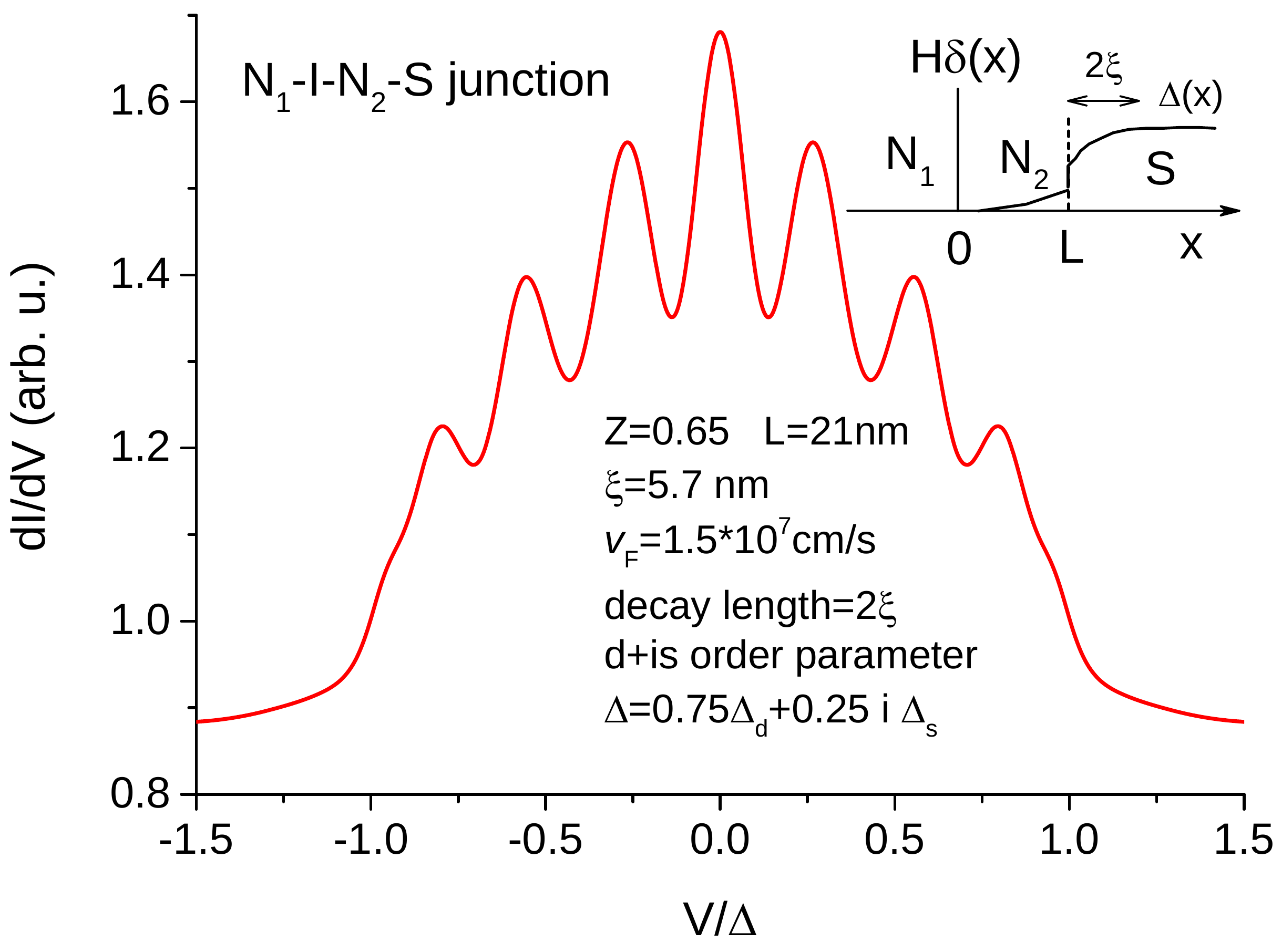}
\vspace{-0mm} \caption{\label{fig:epsart} (Color online) A typical calculated conductance spectrum in the node direction of $\rm N_1-I-N_2-S$ junctions with d+is order parameter comprising a  ZBCP and series of equally spaced bound state peaks. The inset shows a schematic drawing of the model used.}
\end{figure}

\section{Appendix}

Using a potential model shown schematically in the inset of Fig. A4 for an $\rm N_1-I-N_2-S$ junction with d+is order parameter, yields along the node direction the conductance spectrum depicted in Fig. A4 \cite{Lubimova}. In the calculations a modified BTK model \cite{BTK} for a d-wave superconductor is used \cite{N1IN2S}. One can see that the symmetric series of peaks is very similar to the resonances observed in the present study in Fig. 3 (c). These peaks seem to originate in standing waves of Andreev reflections in the N2 region between the $\rm H\delta(x)$ barrier at x=0 and the sharp drop segment at the dashed line of $\rm \Delta(x)$ at x=L. Thus the peaks in Fig. A4 are due to ABSs. We note that MBSs that can also contribute to the ZBCP are not included in this calculation.\\

\newpage

\bibliography{AndDepBib.bib}

\bibliography{apssamp}

\begin{thebibliography}{99}
\label{Bib}

\bibitem{KaneRMP} M. Z. Hasan and C. L. Kane, Colloquium: Topological insulators, Rev. Mod. Phys.  \textbf{82}, 3045 (2010).

\bibitem{Kitaev} A. Yu Kitaev,   Fault-tolerant quantum computation by anyons, Annals of Physics, \textbf{303}, 2 (2003).

\bibitem{Oreg} Y. Oreg, G. Refael, and F. von Oppen, Helical Liquids and Majorana Bound States in Quantum Wires, Phys. Rev. Lett. \textbf{105}, 177002 (2010).

\bibitem{KorenPRB12} G. Koren and T. Kirzhner, Zero-energy bound states in tunneling conductance spectra at the interface of an s -wave superconductor and a topological insulator in NbN/$\rm Bi_2Se_3$/Au thin-film junctions Phys. Rev. B \textbf{86}, 144508 (2012).  arXiv:1207.5352 [cond-mat.supr-con] (2012).

\bibitem{Kouwenhoven} V. Mourik, K. Zuo, S. M. Frolov, S. R. Plissard, E. P. A. M. Bakkers, L. P. Kouwenhoven, Signatures of Majorana fermions in hybrid superconductor-semiconductor nanowire devices, Science \textbf{336}, 1003 (2012).

\bibitem{Moti} A. Das, Y. Ronen, Y. Most, Y. Oreg, M. Heiblum, H. Shtrikman, Zero-bias peaks and splitting in an Al–InAs nanowire topological superconductor as a signature of Majorana fermions, Nature Phys. \textbf{8}, 887 (2012).

\bibitem{KorenEPL13} G. Koren, T. Kirzhner, Y. Kalcheim and O. Millo, Signature of proximity-induced $\rm p_x+ip_y$ triplet pairing in the doped topological insulator $\rm Bi_2Se_3$ by the  s-wave superconductor  NbN, \textit{EPL} \textbf{103}, 67010 (2013).  	 arXiv:1303.0652 [cond-mat.supr-con] (2013).
    
\bibitem{KorenSUST15} Gad Koren, Proximity effects at the interface of a superconductor and a topological insulator in NbN-$\rm Bi_2Se_3$ thin film bilayers, Supercond. Sci. Technol. \textbf{28}, 025003 (2015). arXiv:1409.2975 [cond-mat.supr-con] (2014).

\bibitem{CMarcus}  S. M. Albrecht, A. P. Higginbotham, M. Madsen, F. Kuemmeth, T. S. Jespersen, J. Nyg{\aa}rd, P. Krogstrup and C. M. Marcus, Exponential protection of zero modes in Majorana islands, Nature \textbf{531}, 206 (2016).
    

\bibitem{Butch} N. P. Butch, K. Kirshenbaum, P. Syers, A. B. Sushkov, G. S. Jenkins, H. D. Drew and J. Paglione, Strong surface scattering in ultrahigh-mobility $\rm Bi_2Se_3$ topological insulator crystals, Phys. Rev. B \textbf{81}, 241301(R) (2010).
    
\bibitem{KorenSUST18} G. Koren, Strongly suppressed superconducting proximity effect and ferromagnetism in trilayers of $\rm Bi_2Se_3$ / $\rm SrRuO_3$ / underdoped $\rm YBa_2Cu_3O_x$: A possible new platform for Majorana nano-electronics, arXiv:1802.03728 [cond-mat.supr-con] (2018).

\bibitem{BTK} G. E. Blonder, M. Tinkham, and T. M. Klapwijk, Transition from metallic to tunneling regimes in superconducting microconstrictions: Excess current, charge imbalance, and suppercurrent conversion, Phys. Rev. B \textbf{25}, 4515 (1982).

\bibitem{Tinkham} M. Tinkham, Resistive transition of high-temperature superconductors, Phys. Rev. Lett. \textbf{61}, 1658 (1988).

\bibitem{FuBerg} Liang Fu and E. Berg, Odd-Parity Topological Superconductors: Theory and Application to $\rm Cu_xBi_2Se_3$, Phys. Rev. Lett. \textbf{105},  097001 (2010).


\bibitem{Lubimova} I. Lubimova, Conductance spectra of $\rm N_1-I-N_2-S$ junctions of the cuprates in the ballistic limit, unpublished (2004). See a typical result in the Appendix.

\bibitem{N1IN2S} S. Kashiwaya, Y. Tanaka, M. Koyanagi, H. Takashima and K. Kajimura, Origin of zero-bias conductance peaks in high-$\rm T_c$ superconductors, Phys. Rev. B \textbf{51}, 1350 (1995).




\end{thebibliography}

\end{document}